\documentclass[11pt]{article}
 \usepackage{graphicx}
 \usepackage{epstopdf}
\begin{document}

\begin{center}
\begin{LARGE}
Measuring the Neutron's Mean Square Charge Radius Using Neutron Interferometry\\
\vspace{3ex}
\end{LARGE}

F.~E.~Wietfeldt, M.~Huber\\
{\em Tulane University, New Orleans, LA}\\
\vspace{2ex}
T.~C.~Black\\
{\em University of North Carolina, Wilmington, NC}\\
\vspace{2ex}
H.~Kaiser\\
{\em Indiana University Cyclotron Facility, Bloomington, IN}\\
\vspace{2ex}
M.~Arif, D.~L.~Jacobson, S.~A.~Werner\\
{\em National Institute of Standards and Technology, Gaithersburg, MD}
\end{center}


\begin{abstract}

The neutron is electrically neutral, but its substructure consists of charged quarks so it may have an internal charge distribution. In fact it is known to have a negative mean square charge radius (MSCR), the second moment of the radial charge density. In other words the neutron has a positive core and negative skin. In the first Born approximation the neutron MSCR can be simply related to the neutron-electron scattering length $b_{ne}$. In the past this important quantity has been extracted from the energy dependence of the total transmission cross-section of neutrons on high-Z targets, a very difficult and complicated process. A few years ago S.~A.~Werner proposed a novel approach to measuring $b_{ne}$ from the neutron's dynamical phase shift in a perfect crystal close to the Bragg condition. We are conducting an experiment based on this method at the NIST neutron interferometer which may lead to a five-fold improvement in precision of $b_{ne}$ and hence the neutron MSCR. 
\end{abstract}

\section{Introduction}

The neutron has zero net electric charge, but it is composed of charged quarks so it may have
an internal charge distribution. If so the neutron will interact electrostatically with other charged particles such as
electrons. In a scattering experiment it is useful to parameterize the neutron charge distribution using the Sachs
electric form factor $G_E(Q^2)$,  the spatial Fourier transform of the neutron charge density in the
Breit (brick wall) frame, {\em i.e.} in the frame where the neutron's initial momentum is 
{\boldmath $p$}$_i$ = $-${\boldmath $Q$}$/2$  and its final momentum is {\boldmath $p$}$_f$ = $+${\boldmath $Q$}$/2$ \cite{Sac62}. 
Writing the electric form factor as a multipole expansion in $Q^2$ we have:
\begin{equation}
\label{GE}
G_E(Q^2) = q_n - \frac{1}{6} \langle r_n^2 \rangle Q^2 + \ldots
\end{equation}
At $Q^2 = 0$ the electric form factor is simply the neutron's net charge: $q_n = 0$. In the low $Q^2$ limit $G_E(Q^2)$
is dominated by the second term which contains the neutron mean square charge radius (MSCR) $\langle r_n^2 \rangle$,
the second moment of the neutron charge density:
\begin{equation}
\langle r_n^2 \rangle = \int \rho(r) r^2 d^3r.
\end{equation}
From (\ref{GE}) we see that $\langle r_n^2 \rangle$ determines the slope of the form factor at $Q^2 = 0$:
\begin{equation}
\label{GEslope}
\langle r_n^2 \rangle = -6 \frac{dG_E(Q^2)}{d Q^2} \Big|_{Q^2 = 0} .
\end{equation}
In the subfield of nuclear physics concerned with understanding nucleon structure, one measures $G_E(Q^2)$ as a function of 
$Q^2$ by scattering electrons from light nuclei at various energies and compares it to the predictions of theoretical models. For
this program (\ref{GEslope}) is an important constraint on the shape of $G_E(Q^2)$. 
\par
Theoretically there are several ways of understanding how a nonzero $\langle r_n^2 \rangle$ might arise. The simple SU(6) quark
model predicts $G_E(Q^2)$ to be identically zero, and hence $\langle r_n^2 \rangle = 0$ \cite{Car76}.  
The known $\Delta-n$ mass splitting implies a spin-dependent
quark interaction that breaks SU(6) symmetry. Such a force would also tend to pull the up quark to the center of the neutron and 
push the down quarks to the edge, producing a nonzero $G_E(Q^2)$ and specifically a negative $\langle r_n^2 \rangle$ \cite{Isg81}.
In another picture every neutron spends part of its time as a 
virtual proton-pion pair \cite{Fer47}. In this state the lighter $\pi^-$ orbits the heavier proton, making a negative
$\langle r_n^2 \rangle$. Recent theoretical efforts to calculate $\langle r_n^2 \rangle$ produce a negative sign
but a large uncertainty on its magnitude \cite{Car00,Kar02,Ara03}.
\par
In an experiment, the total coherent scattering length for neutron-atom scattering 
(far from any resonances and neglecting small effects such as
the neutron's electric polarizability) is given by:
\begin{equation}
\label{E:bCoh}
b_{\rm coh} = b_N + Z \left[ 1 - f(\mbox{\boldmath $q$}) \right] b_{ne}
\end{equation}
where $b_N$ is the nuclear scattering length and $b_{ne}$ is the neutron-electron scattering length, responsible for the electrostatic
interaction between the neutron and a charged particle ({\em e.g.} electron). The first term in brackets corresponds to the nuclear electric charge and the second term to the atomic electrons. The form factor $f(\mbox{\boldmath $q$})$ is the spatial Fourier transform of the atomic electron charge distribution for momentum transfer $\mbox{\boldmath $q$}$.
\par
The net charge of the particle and its internal charge distribution both contribute to $b_{ne}$. For neutron-atom scattering at low momentum transfer the second term in (\ref{GE}) dominates. From scattering theory, using the 1$^{\rm st}$ Born approximation,
one obtains:
\begin{equation}
\langle r_n^2 \rangle = 3\,a_0 \left( \frac{m_e}{m_n} \right) b_{ne} = \mbox{ (86.34 fm)$\cdot b_{ne}$}.
\end{equation}
In recent experimental determinations of $b_{ne}$, total transmission
cross sections are measured for epithermal and thermal neutrons on lead and bismuth targets \cite{Ale86,Koe95,Kop97}. 
These studies, and also
the earlier work of Krohn and Ringo \cite{Kro73} using noble gas targets, contribute to the Particle Data Group's recommended value of
$\langle r_n^2 \rangle = -0.1161 \pm 0.0022$ fm$^2$ \cite{PDG}. This includes a scale factor of 1.3 to account for a minor disagreement.
There is a much larger and still unexplained disagreement between measurements of $b_{ne}$ from transmission 
measurements and older determinations neutron diffraction on crystals \cite{Ale75}.

\section{A New Method}
\begin{figure}
\begin{center}
\includegraphics{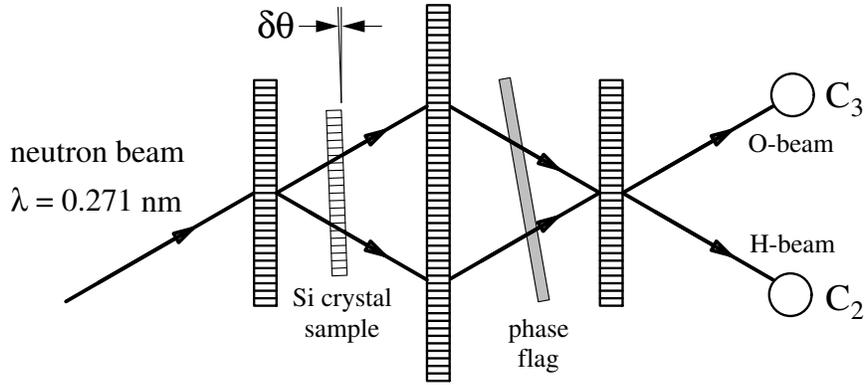}
\caption{\label{F:setup} A sketch of the experimental setup.}
\end{center}
\end{figure}

Recently S.~A.~Werner proposed a new approach for a precision measurement of $b_{ne}$ using neutron interferometry. The basic experimental setup is illustrated in Figure \ref{F:setup}, which shows a standard LLL perfect crystal silicon neutron interferometer and the usual phase flag ({\em e.g.} fused silica). The neutron beam is split into two paths by Bragg diffraction, and the relative phase shift caused by different path lengths in the phase flag produces interference fringes in the neutron counters C$_2$ and
C$_3$. Detailed introductions to the theory and operation of neutron inteferometers can be found in \cite{Sears,KR,RW}.
A separate perfect crystal silicon sample, cut from the same ingot as the interferometer, is placed
across both downstream paths. This sample can be rotated about all three axes. It is aligned precisely to the Bragg condition
(parallel to the interferometer blades) and then rotated by a small misset angle $\delta\theta$ as indicated. The neutron phase
as a function of $\delta\theta$ over a range of $\pm10$ arcsec from Bragg is then measured.
\par
\begin{figure}
\begin{center}
\includegraphics[width=4.0in]{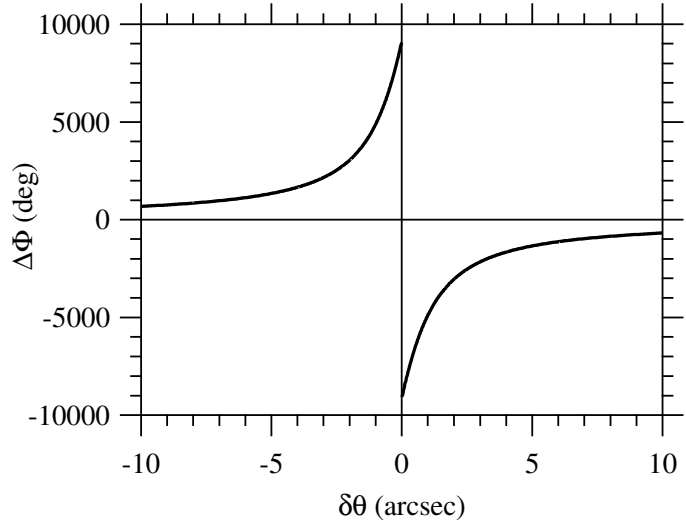}
\caption{\label{F:delPhi} The theoretical relative phase $\Delta\Phi$ as a function of the misset angle $\delta\theta$ from
Bragg, calculated from Equation \ref{E:delPhi}. }
\end{center}
\end{figure}

Very close to Bragg the neutron obtains a {\em dynamical} phase shift that originates from the transition to the $\alpha$ and $\beta$ solutions for the internal wave vector. According to dynamical diffraction theory this is given by \cite{RW2}:
\begin{equation}
\Phi_{\rm dyn} = \frac{\nu_H}{\cos\theta_B} \left( -y \pm \sqrt{1 + y^2} \right) D
\end{equation}
where the minus (plus) sign corresponds to the $\alpha$ ($\beta$) solution. Here $\nu_H$ is the probability for Bragg scattering
per unit path length, related to the crystal structure function $F_H$ for this reflection by 
$\nu_H = F_H \lambda / V_{\rm cell}$.
The parameter $y$ is a scaled misset angle:
\begin{equation}
y = \frac{k \sin\theta_B}{2\, \nu_H} \delta\theta,
\end{equation}
$D$ is the crystal thickness, and $\theta_B$ is the Bragg angle. Because the crystal sample extends through both beam paths the kinematic phase cancels and the dynamical phase adds, producing a net relative phase:
\begin{equation}
\label{E:delPhi}
\Delta\Phi = \frac{\nu_H}{\cos\theta_B} \left( y \pm \sqrt{1 + y^2} \right) D
\end{equation}
with the plus (minus) sign corresponding to positive (negative) misset angles. A plot of $\Delta\Phi$ {\em vs.} $\delta\theta$ is
shown in Figure \ref{F:delPhi}. 
\par
The sample crystal is suspended on a shaft connected to a 10 cm lever, which is moved by a self-calibrated PZT nanopositioner
that steps over a range of 25 $\mu$m with 0.1 nm precision (P-753, Physik Instrumente, Karlsruhe, Germany \cite{NISTdisc}). 
The shaft rotates on a nylon flexure bearing, there are no sliding or rolling contacts, so the crystal rotates smoothly and the
misset angle can be measured to 10$^{-4}$ arcsec.
\par
The neutron interferometer can measure the net relative phase shift $\Delta\Phi$ to within 1$^{\circ}$ in a single interferogram (in about 1000 s) The silicon lattice constant, neutron wavelength, and crystal thickness $D$ are all known at the 10$^{-5}$ precision level. Therefore by fitting the measured phase shift as a function of misset angle to Equation \ref{E:delPhi} one can with some effort determine $\nu_H$, and hence $F_H$, to an absolute precision of about $3\times10^{-5}$. Near the Bragg condition the structure factor is related to the coherent scattering length (see \ref{E:bCoh}) by:
\begin{equation}
\label{E:onB}
F_H = \sqrt{32} \left( b_N + Z \left[ 1 - f(\mbox{\boldmath $H$}) \right] b_{ne} \right).
\end{equation}
In our case $f(\mbox{\boldmath $H$}) = 0.7526$ \cite{Cro68} is the form factor calculated for the reciprocal lattice vector \mbox{\boldmath $H$}
corresponding to the (111) reflection in silicon. In Equation \ref{E:onB}, $b_N$ is about 240 times larger than
$Z \left[ 1 - f(\mbox{\boldmath $H$}) \right] b_{ne}$. By rotating the crystal off Bragg far enough that the dynamical phase is negligible
we can measure $b_N$ {\em in situ} with a relative uncertainty of less than 10$^{-5}$, limited only by the uncertainty in the phase measurement, and subtract it from Equation \ref{E:onB}. This will allow a determination of $b_{ne}$ and hence 
$\langle r_n^2 \rangle$ with a net uncertainty of about 0.7\%, more than a factor of three improvement from the present accepted value,
and using a technique systematically much different from other precision measurements.
If the experiment with silicon is successful we may repeat it using a perfect germanium crystal, improving the sensitivity
by the increase in $Z$: a factor of 2.3. This may lead to a 0.3\% measurement of $b_{ne}$ and $\langle r_n^2 \rangle$, more than a factor
of five improvement.
\par
We note that this method is related to the experiment of Shull which first demonstrated Pendell\"{o}sung interference fringes in neutron dynamical diffraction  \cite{Shu68}. The structure factor $F_H$  was precisely determined
from the fringe separation and $b_{\rm coh}$ was obtained for the (111) reflection in silicon via Equation \ref{E:onB}. The main advantages of our approach are (1) an improved overall precision due to the interferometric measurement of the neutron phase and (2) the ability to make a relative subraction of $b_N$ from Equation \ref{E:onB} {\em in situ}.


\begin{thebibliography}{99}
\bibitem{NISTdisc} Certain trade names and company products are mentioned in the text or identified in illustrations in order to adequately specify the experimental procedure and equipment used. In no case does such identification imply recommendation or endorsement by the National Institute of Standards and Technology, nor does it imply that the products are necessarily the best available for the purpose.
\end{thebibliography}
\end{document}